\documentclass[fleqn,usenatbib]{mnras}

\usepackage{mathptmx}

\usepackage[T1]{fontenc}
\usepackage{ae,aecompl}

\usepackage{graphicx}	

\title[Multiple populations in M~4 AGB stars]{Multiple populations along the asymptotic giant branch of the globular cluster M~4}  

\author[C.~Lardo et al.] {
C.~Lardo,$^{1}$\thanks{E-mail: c.lardo@ljmu.ac.uk}
M.~Salaris,$^{1}$
A.~Savino,$^{1,2}$
P.~Donati,$^{3,4}$
P.~B.~Stetson,$^{5}$
 and S.~Cassisi$^{6}$
 \\
$^{1}$  Astrophysics Research Institute,  Liverpool John Moores University, IC2, Liverpool Science Park, 146 Brownlow Hill, Liverpool L3 5RF, UK\\
$^{2}$ Kapteyn Astronomical Institute, University of Groningen, Postbus 800, 9700 AV Groningen, The Netherlands \\
$^{3}$ INAF --Osservatorio Astronomico di Bologna, via Ranzani 1, I-40127 Bologna, Italy\\
$^{4}$ Dipartimento di Fisica e Astronomia, via Ranzani 1, I-40127 Bologna, Italy\\
$^{5}$ Herzberg Astronomy and Astrophysics, National Research Council Canada, 5071 West Saanich Road, Victoria, BC V9E 2E7, Canada\\
$^{6}$ INAF-Osservatorio Astronomico di Teramo, Via M. Maggini, I-64100 Teramo, Italy
}

\date{Accepted 2016 December 21. Received 2016 December 21; in original form 2016 September 23}

\pubyear{2017}

\begin{document}
\label{firstpage}
\pagerange{\pageref{firstpage}--\pageref{lastpage}}
\maketitle

 \begin{abstract}
Nearly all Galactic globular clusters host stars that display characteristic abundance anti-correlations, like 
the O-rich/Na-poor pattern typical of field halo stars, together with O-poor/Na-rich additional components.
A recent spectroscopic investigation questioned the presence of O-poor/Na-rich stars amongst a sample of asymptotic giant branch 
stars in the cluster M~4, at variance with the spectroscopic detection of a O-poor/Na-rich component along both the cluster red giant branch and 
horizontal branch.
This is contrary to what is expected from the cluster horizontal branch morphology and horizontal branch stellar evolution models. 
Here we have investigated this issue by  
employing the $C_{UBI}=(U-B)-(B-I)$ index, that previous studies have demonstrated to be very effective 
in separating multiple populations along both the red giant and asymptotic giant branch sequences. 
We confirm previous results that the RGB is intrinsically broad in the $V-C_{UBI}$ diagram, with the presence of two components which nicely 
correspond to the two populations identified by high-resolution spectroscopy. We find that AGB stars 
are  distributed over a wide range of $C_{UBI}$ values, in close analogy with what is observed for the RGB, 
demonstrating that the AGB of M~4 also hosts multiple stellar populations.
\end{abstract}

\begin{keywords}
globular clusters: general -- globular clusters:individual: M~4 -- stars: abundances -- 
stars: Hertzsprung-Russell and colour-magnitude diagrams -- stars: AGB and post-AGB
\end{keywords}


\section{Introduction}
\label{intro}

During the course of the last fifteen years it has become increasingly well established 
that individual Galactic (and extragalactic) globular clusters host a stellar {\sl first population} (FP) with initial chemical abundance ratios similar 
to the ones of field halo stars, plus additional sub-populations (that we collectively denote as second population --SP), 
each one characterised by its
own specific variation of He (increased), C (decreased), N (increased), O (decreased), Na (increased), and sometimes
Mg (decreased) and Al (increased) abundances \citep[see, e.g.,][]{gsc}, compared to FP ratios.
These abundance patterns give origin to well defined (anti-)correlations
between pairs of light elements, the most characteristic ones 
being the Na-O and the C-N anti-correlations. 

According to the currently most accepted scenarios,   
the observed abundance patterns are produced by high temperature proton captures either at the bottom of the
convective envelope of massive asymptotic giant branch (AGB) stars \citep[see, e.g.,][]{agb}, or in 
main sequence fast rotating massive stars \citep{frms},   
or supermassive stars \citep{d2015}, belonging to the cluster FP.
This chemically processed material is transported to the surface either by convection (in AGB stars and fully convective supermassive stars) 
or rotational mixing (in fast rotating massive stars), and spread in the intra-cluster medium by stellar winds. 
SP stars are supposed to have formed out of this gas, with a time delay  
that depends on the type of polluter, but it is always very short compared to the typical age of these old stellar systems.

In recent years some controversy has arisen about the presence of SP stars along the AGB of some Galactic globular cluster (GGC).
Standard stellar evolution \citep[see, e.g.,][]{cs13} dictates that 
horizontal branch (HB) stars with masses below $\sim 0.50-0.53{M_{\odot }}$ (the exact value depending on 
the initial chemical composition) fail to reach the AGB phase \citep[the so-called {\sl AGB-manqu\'e} stars, see][]{gr},  
and SP stars are indeed expected to have on average a lower mass along the HB with 
respect to FP stars. In fact, given that SP stars are 
typically (to a more or lesser degree) He-enhanced, their HB progeny originate from RGB stars with a lower mass compared to 
the FP RGB population, hence will be on average less massive (and bluer) 
than FP HB stars, if the RGB mass loss is approximately the same for both FP and SP objects. 
The AGB of GGCs with a blue HB may therefore lack at least a fraction of SP stars, 
compared to what is seen along the RGB. 

\citet{gh15} found SP stars along the AGB of the GGCs 
M13, M5, M3, and M2, by combining the $H$-band Al abundances obtained 
by the Apache Point Observatory Galactic Evolution Experiment (APOGEE) survey with ground-based optical photometry, although 
they didn't study whether the SP/FP ratio was consistent with the corresponding ratio found on the RGB and the HB morphology.

The spectroscopic study by \citet{campbell} found no SP stars along the AGB of NGC6752 (a cluster with a moderately extended blue HB), 
at odds with results from detailed synthetic HB modelling  
\citep{csp} showing that, for the observed HB morphology, a SP component should be present along the AGB, 
that should lack only the more extreme Na-enhanced population.
A later spectroscopic analysis by 
\citet{ll} found AGB SP stars with moderate Na enhancement, as predicted by synthetic HB models.

Very recently \citet{maclean} have found spectroscopically a lack of SP stars in M~4.
These authors claim that the AGB abundance distribution is consistent with all AGB objects being FP stars, although their discussion  
mentions that, due to the errors on [Na/Fe] and [O/Fe] abundance ratios, the presence of a few SP AGB stars cannot be excluded.
This cluster has [Fe/H]$\sim -$1.1 \citep[see, e.g.,][and references therein]{mvm} and a HB that does not extend above $\sim$9000~K,  hence 
it does not show any blue tail in the $BV$ colour-magnitude-diagram (CMD), and --as we will see in Sect.~\ref{analysis}-- 
is populated by objects with masses above $0.55{M}_{\odot }$. HB stellar models with the cluster chemical composition and this minimum  
mass do move to the AGB after core-He exhaustion. Moreover, the spectroscopic studies by \citet{mvm} and \citet{villanova} have targeted some of 
the bluest HB stars in this cluster, and have measured Na abundances roughly as high as the highest Na abundances measured along the RGB 
\citep[][have also observed stars belonging to the red portion of the HB, that turned out to have typical FP composition]{mvm}.
Given that these bluest Na-rich HB stars are expected to evolve on the AGB, a lack of SP stars along this phase
is totally unexplainable by stellar evolution.
Notice that the bluest HB stars are cooler than the observed $T_{\rm{eff}}$ threshold for the onset of radiative levitation 
\citep[$\sim$11000-12000~K, see, e.g.,][]{grundahl,b03,b16}. This means that one cannot even consider 
the hypothesis by \citet{campbell} --invoked to explain their 
results for NGC6752-- that enhanced HB mass loss associated to the surface metal enhancement caused by radiative levitation may push 
HB stars into the {\sl AGB-manqu\'e} regime \citep[but see also][for a detailed investigation on this issue]{vink}.

Given the uncertain claims of \citet{maclean} paper, we revisit the issue of the AGB population in M~4, 
using photometry. Recent observational and theoretical analyses have 
in fact shown how appropriate combinations of broad- and/or intermediate-band filters are 
capable to reveal the presence of multiple populations along the CMD branches of 
GGCs \citep[see, e.g.,][and references therein]{marino08, yong08, sbordone, m15a, p15}
Here we make use of $UBVI$ magnitudes and the $C_{UBI}=(U-B)-(B-I)$ index \citep[similar to the $(U-B+I)$ index by][]{milone47tuc}, that 
has been successfully employed to reveal the  
presence multiple stellar populations along the AGB in a number of GCs \citep[see e.g.][]{milone6752, sumo, miloneM2, milone2808, 
nardiello15, gh15}.

The next section describes briefly the photometric data, followed by an analysis of the cluster AGB and RGB using $V-C_{UBI}$ 
diagrams, and conclusions.

\section{Photometric data}
\label{data}

Our optical observations of M~4 are the same presented in \citet{sumo} and \citet{stetson14}.
The \citet{stetson14} dataset was further complemented with the observations listed in Table~\ref{logobs}, for a grand 
total of 5,351 individual CCD images obtained during the course of 25 observing runs.
We refer to \citet{stetson14} for a description of the observational material.

\begin{table*}

\setlength{\tabcolsep}{0.28cm}

\caption{Logs of observations in optical bands used in this paper, which complement the dataset presented in \citet{stetson14}}
\label{logobs}
\begin{tabular}{@{}lllllcccccc}
\hline
Run &  ID      &        Dates     & Telescope  &    Camera   &    $U$  & $B$  & $V$  & $R$ &  $I$  & Multiplex  \\
          \hline
19$^{a}$  & wfi42        &   2000 Apr 01               &    ESO/MPI 2.2m   &   WFI                & --   &3   &4  &--&   4 &   8\\
20$^{b}$  & wfi40        &  2007 Jul 09-13            &    ESO/MPI 2.2m   &   WFI                & -- & -- & 10 & --  &11 &   8\\
21$^{c}$  &Y1008       &     2010 Aug 14-15        &    CTIO 1.0m         &  Y4KCam ITL SN3671 &  -- & 11  &12  &-- & -- &   -\\
22$^{d}$   &efosc1305 & 2013 May 14-15           &    ESO NTT 3.6m   &   EFOSC/1.57 LORAL &   -- & --  &--  &--  &14 &   -\\
23$^{e}$   &lcogt4        &     2014 Feb 10             &    Sutherland 1.0m  &  CCD                & --  &--  &10 & --  &10&    -\\
24$^{f}$  & lcogt1        &     2014 Mar 27, Apr 10 &   Sutherland 1.0m   & CCD                 &--  &-- & 18  &-- &  9 &   -\\
25$^{g}$  & efosc1406 &  2014 Jun 27                 &     ESO NTT 3.6m   &  EFOSC/1.57 LORAL    &-- & --  &13  &--   &9 &   -\\

\hline
\end{tabular}
\medskip

\flushleft{{\bf Notes:} The column
"multiplex" refers to the number of individual CCD chips in the
particular camera, which were treated as independent detectors. ($a$) ESO Program ID 164.O-0561(F); ($b$)  ESO Program ID 079.D-0918(A); ($c$)  SMARTS Project ID Sejong10B; ($d$)  ESO Program ID 091.D-0711(A); ($e$)  Proposal ID STANET-002; ($f$) Proposal ID STANET-002; ($g$)  ESO Program ID 093.D-0264(A).\\
}
\end{table*}

All the CCD images were reduced and analysed by Peter~B.~Stetson using the
{\tt DAOPHOT/ALLSTAR/ALLFRAME} suite of programs. These
data were then calibrated to the Johnson $UBV$ , Kron/Cousins $RI$
photometric system of \citet{stetson00,stetson05}
\footnote{Please see \url{http://www1.cadc-ccda.hia-iha.nrc-cnrc.gc.ca/community/STETSON/standards/}; 
see also \url{http://www.cadc.hia-iha.nrc.gc.ca/community/STETSON/
homogeneous/archive/}}.
This analysis
resulted in 80480 stars in the M~4 field with calibrated photometry
in $V$ and at least one of $B$ or $I$; 79146 stars had
calibrated photometry in all three of $B$, $V$ , and $I$; and
31791 of them had calibrated photometry in $U$,
$B$, and $I$.
Further details on the reduction and analysis of the photometric material 
used in this paper can be found in \citet{stetson14}.

Given the large value of $E(B-V)$ ($E(B-V)=0.35-0.40$) 
and an anomalous large $R_V=A_V/E(B-V)\sim$3.8 in the line of sight 
towards M~4 \citep[see, e.g.,][and references therein]{hendricks, kaluzny}, 
we have corrected for the possible presence of differential reddening (DR) 
across the face of the cluster in the field of view of our observations.
We followed the recipe described in
\citet{donati14}, that revises the method by \citet{m12}. The reader is referred to the 
former paper for details. In brief, we have first considered main sequence (MS) 
stars in the $V-(B-V)$ CMD and drawn a reference fiducial line through the middle of the observed MS 
between $V=17.5$ and $V$=20.0. For each star in the photometry we then selected about 
30 of the nearest (spatially) MS objects and averaged their distance along the reddening vector 
direction from the MS ridge line in the CMD. This value was adopted as a
DR correction for each star. 
Notice that the mean value of the DR correction is not necessarily equal to zero, for it depends on the colour 
distribution of the MS stars around the adopted reference ridge line.

To correct the individual magnitudes for DR we adopted $R_V=3.8$, 
and $A_U=1.43A_V$, $A_B=1.22A_V$, $A_I=0.62A_V$, as 
obtained by convolving the \citet{ccm} reddening law with an ATLAS9 $\alpha$-enhanced ([$\alpha$/Fe]=0.4) spectrum 
\citep{bastibc} for [Fe/H]=$-$1.0, log($g$)=4.5 and $T_{\rm{eff}}=5800$~K, corresponding to about 1~mag 
below the turn off in $V$ for a 12.5~Gyr isochrone with the same chemical composition (see Sect.~\ref{analysis})

We find that the total range of reddenings $\Delta E(B-V)$ spanned by 90\% of the cluster stars is equal to $\sim$0.10~mag, 
consistent with the reddening map presented in \citet{sumo}.


%

\begin{figure}
\centering
\includegraphics[width=0.85\columnwidth]{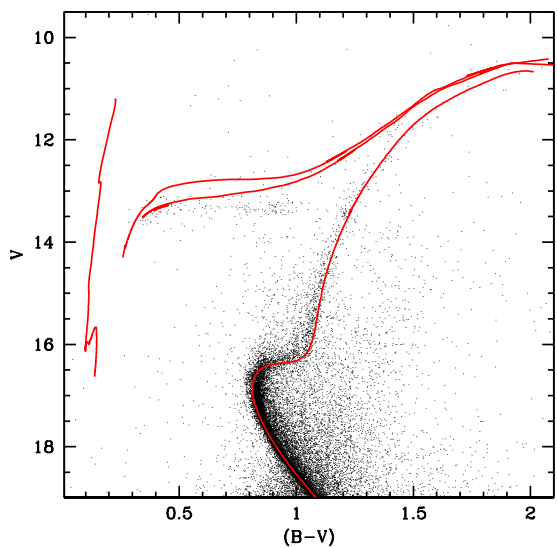}
\caption{Comparison of a 12.5~Gyr isochrone with [Fe/H]= --1.0 and [$\alpha$/Fe]= 0.40, with the 
the cluster $V-(B-V)$ CMD. We display also HB evolutionary tracks for masses equal to (from the red to the blue) 
0.58, 0.56 and 0.495~$M_{\odot}$, respectively. A distance modulus 
$(m-M)_0$= 11.30, and reddening $E(B-V)$= 0.34 ($R_V$= 3.8 -- see text for details) have been employed.}
\label{cmdiso}
\end{figure}

A word of caution about the DR corrections determined in this fashion is in general necessary. 
According to the derivative $\Delta Y/ \Delta (B-V) \sim$1 for MS stars 
\citep[obtained from][models at varying $Y$]{basti}, an $Y$ range amongst the cluster sub-populations 
produces an intrinsic width of the MS, that could potentially cause an overestimate of the range spanned by E($B-V$). This potential systematic error 
will depend on the range $\Delta Y$ and the relative fraction of FP and SP MS stars employed to determine 
the DR correction for each object in the photometry. 

As pointed out by our referee, this is however not a matter of concern for M~4. 
Determinations of $\Delta Y$ in the literature range 
from negligible values \citep{valc13}, up to just $\sim$0.02 \citep{villanova, nardiello15}. 
The fraction of SP stars in this cluster is approximately the same between the central \citep{milone16} 
and external \citep{marino08, nardiello15} regions analyzed so far, being equal to 30-40 \%; this fraction, together with a maximum $\Delta Y=$0.02 
would cause systematic effects on the reddening below 0.01~mag.

\section{Analysis}
\label{analysis}

Before starting the analysis of M~4 $V-C_{UBI}$ diagram, we display in Fig.~\ref{cmdiso} 
the $V-(B-V)$ CMD corrected for DR, compared to theoretical isochrones. 
We employ here BaSTI \citep{basti} $\alpha$-enhanced isochrones \footnote{\url{http://www.oa-teramo.inaf.it/BASTI}}
with [Fe/H]=$-$1.01 ([$\alpha$/Fe]=0.4), He mass fraction $Y$=0.251, a distance modulus $(m-M)_0$=11.30 as obtained by  
\citet{kaluzny} from the analysis of three cluster eclipsing binaries, $E(B-V)$=0.34 and $R_V$=3.8.
Notice that in the $BV$ CMD theoretical isochrones with a standard $\alpha$-enhanced metal mixture 
(employed in both the stellar model and stellar spectra calculations) are appropriate to match also SP stars 
\citep[see][]{sbordone, cmpsf}.
Also, the range of He abundance (expressed as mass fraction $Y$) 
estimated for the cluster populations is very small, with $\Delta Y$ equal to 0.02 or less  
\citep{villanova, valc13, nardiello15}.

The purpose of this comparison --that is not meant to be a perfect fit model-theory--  is to show very clearly 
how HB stellar models predict that all HB stars in this cluster evolve to the AGB phase. The mass of the 
the HB track that starts the evolution at the hottest end of the observed HB is equal to $\sim$0.58$~M_{\odot}$. 
Observationally the onset of radiative levitation is at $\sim$12000~K, corresponding to the zero age HB location of 
the 0.56$~M_{\odot}$ track shown in the figure. Notice how this threshold (in the hypothesis it could trigger 
a very strong wind that decreases rapidly the stellar mass) is well beyond the hot end of the observed HB.
We also display, just for the sake of comparison, the most massive (M=0.495~$M_{\odot}$) {\sl AGB-manqu\'e} star.

As shown by \citet{marino08} and \citet{sbordone} the $(U-B)$ colour is very sensitive to light-element variations, whereas the $(B-I)$ colour 
is unaffected and is mainly sensitive to $T_{\rm{eff}}$ variations. 
The combination $C_{UBI}=(U-B)-(B-I)$ is therefore able to segregate FP and SP stars in a $V-C_{UBI}$ diagram as discussed in \citet{milone6752}, 
\citet{sumo} and \citet{gh15}, producing 
either multimodal or very broad sequences, broader than expected from the photometric error. This is true also for AGB stars, as shown empirically by \citet{gh15}, 
\citet{sumo}, \citet{miloneM2,milone2808}, and \citet{nardiello15}.

\begin{figure}
\centering
\includegraphics[width=0.99\columnwidth]{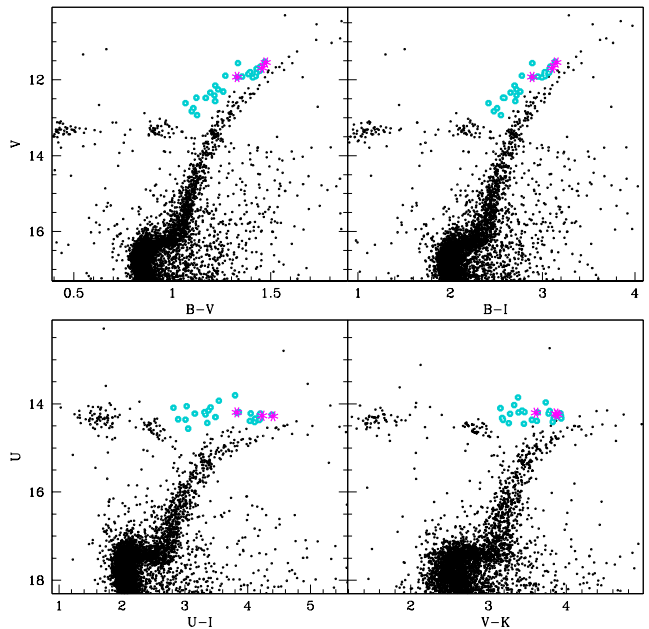}
\caption{Selection of AGB star candidates in different CMDs. From top to bottom: $V, B-V$; $V, B-I$; $U, U-I$.
Empty cyan symbols are AGB stars. Magenta asterisks mark candidate AGB stars in common with \citet{marino08}.}
\label{CLASS}
\end{figure}

To look for signatures of SP stars along M~4 AGB stars in the $V$-$C_{UBI}$ plane, we follow a standard procedure similar to the one outlined in \citet{sumo}. 
We firstly cleaned the CMD from foreground/background contamination, i.e. cluster membership has been assigned on the basis of the 
source position in the $(B-I)$ versus $(U-V)$ plane \citep[see Fig.~2 in][]{sumo}.  As a first step, we defined AGBs as shown in the top left panel of Figure~\ref{CLASS}. Then, we plotted the selected stars in other CMDs which provide the cleanest separation of the evolved sequences to help verify the identification (see Figure~\ref{CLASS}). To this aim, we cross-correlated  our $UBVI$ photometry with the 2MASS catalogue\footnote{ \url{http://www.ipac.caltech.edu/2mass/}.} to carefully inspect the position of selected AGB stars in CMDs including near-infrared colours.
All selected stars are compatible with being AGBs in all diagrams displayed in Figure~\ref{CLASS}.
This demonstrates that they cannot be RGBs migrated to the AGB sequence due to photometric errors. Indeed, no random errors can explain that the selected stars are always located in a well defined sequence bluer of the main body of the RGB in different CMDs obtained from independent measurements/datasets.
From all the above we conclude that the selected stars are indeed AGBs. 

We finally notice that the observed AGB/RGB ratio  (R$_{\rm {AGB/RGB}}$= 0.44 $\pm$ 0.23) agrees 
within the errors with the theoretical
one  (R$_{\rm {AGB/RGB}}$= 0.37). This by itself is obviously not a proof that all candidate 
AGBs are indeed AGB stars, but just a consistency check
for our empirically derived value of R$_{\rm {AGB/RGB}}$.

To assess whether some significant residual contamination may still be present in our AGB sample after the field-star subtraction, 
we used the code {\tt TRILEGAL} \citep{girardi05}. 
We determined the fraction of Galactic field stars within a box $\simeq$30\arcmin$\times$30$\arcmin$ 
(i.e. comparable to the FoV of our photometric catalogue) centred on the cluster centre that survive our cluster AGB selection. 
We found one star with colours and magnitudes   
consistent with being on the cluster AGB, to be compared with the two AGB objects rejected with our field-subtraction procedure.
This implies that any residual field contamination is very likely negligible.

\begin{figure}
\centering
\includegraphics[width=0.90\columnwidth]{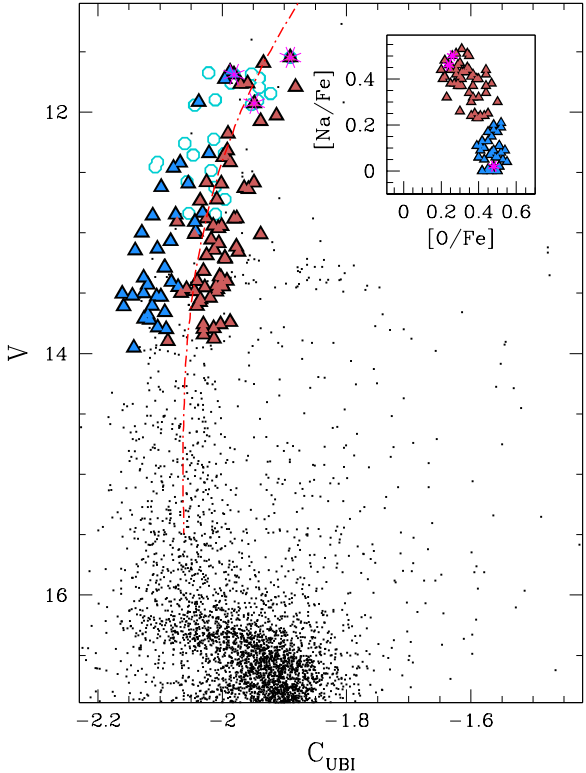}
\caption{$V-C_{UBI}$ diagram is shown. AGB stars and stars in common with the \citet{marino08} study are marked with the same symbols 
as in Figure~\ref{CLASS}.  
We display also a fiducial reference line (dash-dotted line) employed to quantify the 
broadening of the RGB and AGB sequences (see text for details). This line is just for reference, and is not meant to separate in this plot the location 
of FP and SP stars. 
The inset shows the NaO anti-correlation found by 
\citet{marino08} in M~4 RGB stars. FP and SP  stars are plotted as blue and red triangles respectively in the $V-\Delta$$C_{UBI}$ plane. }
\label{cubi}
\end{figure}

\begin{figure}
\centering
\includegraphics[width=0.95\columnwidth]{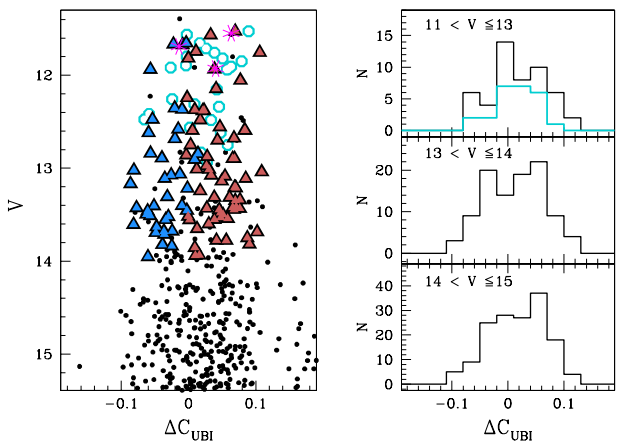}
\caption{{\em Verticalised} $V-\Delta$$C_{UBI}$ diagram for bright giant stars (see text for details). The number distribution of AGB (cyan histogram) and RGB (black histogram). $\Delta C_{UBI}$ index differences  with respect to the RGB fiducial in three $V$-magnitude bins are also shown. Magenta asterisks mark candidate AGB stars in common with \citet{marino08}.
}
\label{histograms}
\end{figure}

Fig.~\ref{cubi} shows the $V-C_{UBI}$ diagram, with AGB stars highlighted. As already shown by \citet{sumo}, the RGB shows a large $C_{UBI}$ spread at fixed $V$, that exceeds the spread due 
to photometric errors, and a hint of bimodality can also be observed.

 Detailed [Na/Fe] and [O/Fe] abundances do exist for a number of bright RGB stars in this cluster 
\citep[e.g.][]{marino08,carretta09}, and in the inset of Fig.~\ref{cubi} we show the NaO anti-correlation  measured by \citet{marino08}. The distribution of RGB stars in the [Na/Fe]-[O/Fe] plane is 
clearly bimodal, with FP stars being clustered around [Na/Fe]$\simeq$0.1 and [O/Fe]$\simeq$0.5 dex. Conversely SP stars show enhanced Na abundances and  are depleted in their O content. FP and SP 
stars --as defined by spectroscopy-- appear clearly segregated along two parallel sequences in the $V-C_{UBI}$ diagram (see Fig.~\ref{cubi}). This shows how the $C_{UBI}$ index is very effective in 
separating the different stellar populations hosted in M~4 \citep[][]{sumo}. AGB stars at a given $V$ span a very similar $C_{UBI}$ range as RGB stars, suggesting that multiple stellar population are 
present also along the AGB stage. 

A reference fiducial line in the $V-C_{UBI}$  diagram (shown Fig.~\ref{cubi}) has been used to {\em verticalise} the $V-\Delta C_{UBI}$ diagram and 
quantify the spread of both AGB and RGB sequences.  
$\Delta C_{UBI}$ differences have been derived by subtracting 
from the $C_{UBI}$ index of each star the corresponding value of the fiducial at the same $V$ magnitude \citep[e.g.][]{lardo11}, and the resulting diagram is shown in   
the top panel of Fig.~\ref{histograms}, where spectroscopic FP and SP RGB stars, and AGB stars are marked. 
The lower panels of Fig.~\ref{histograms} display the number distributions of RGB and AGB $\Delta C_{UBI}$ values 
in three $V$-magnitude ranges. 
We find that AGB stars span the same total $\Delta C_{UBI}$ range as the RGB ones in the common magnitude range (11 $< V \leq$ 13), 
and share the same 1$\sigma$ spread (0.04~mag).  We adopt the bootstrapped $\sigma$ as the uncertainty on our estimate of the observed spread
and find that the spread in colour in $\Delta$ C$_{UBI}$ has standard deviation equal to  0.040 $\pm$ 0.005 and 0.041 $\pm$ 0.005 for AGBs and RGBs in the magnitude range 11$<  V \leq$ 13 mag, respectively. We also determined with a Kolmogorov-Smirnov (KS) test 
that the probability we can reject that hypothesis that the observed $\Delta C_{UBI}$ distributions of AGB and RGB stars (11 $<$ V $\leq$ 13) 
are the same, is well below the standard 95\% threshold (it is equal to just 50\%). Finally, we performed a  KS test on the distribution of the bootstrapped estimates of both the mean ($\Delta$ C$_{UBI}$) and the dispersion of $\Delta$ C$_{UBI}$ for the distributions of AGB and blue and red RGB stars brighter than $V=$ 13 mag. We find that both the mean and the dispersion of $\Delta$ C$_{UBI}$ for AGB and (blue+red) RGB  stars are compatible with being the same, while the same statistical test shows that the mean and the dispersion of $\Delta$ C$_{UBI}$ for AGB, blue  RGB, and red RGB star groups (where the blue and red RGB stars are taken into account separately) cannot be extracted from the same parent distribution.
This leads us to conclude that the distribution in colours of AGB and RGB stars in the C$_{UBI}$ index are statistically indistinguishable.

The demonstrated ability of the $C_{UBI}$ index to separate FP and SP stars along both RGB and AGB \citep[e.g.][]{sumo,gh15,miloneM2,milone2808,nardiello15}, plus the broad and consistent  
distribution of $C_{UBI}$ covered by both RGB (where the presence of FP and SP stars is established spectroscopically) and AGB stars 
make a compelling case for the presence of SP stars also along the AGB of M~4.
These conclusions do not depend on the DR correction we performed, as we recovered the same result 
using as input catalog the M~4 photometry where DR variations have not been taken into account. 

Also, Figures~\ref{cubi} and ~\ref{histograms} show three of the stars observed by \citet{marino08} that should be classified as AGB rather than RGB stars 
according to their CMD location, as shown in Fig.~\ref{CLASS}. 
Based on their Na abundances, two of them are SP stars, while the remaining one has both [Na/Fe] and [O/Fe] abundances as the field at the same metallicity 
(see inset in Figure~\ref{cubi}). This further demonstrates that multiple population can be found also among AGB stars.

\section{Conclusions}
\label{summary}
We have presented a multi-wavelength photometric analysis of RGB and AGB stars in the GGC M~4, to establish whether SP stars are present along the cluster AGB sequence, as expected 
from the observed HB morphology, results from HB star spectroscopy, and stellar evolution calculations. 
The analysis of the $C_{UBI}$ index distribution of AGB and RGB stars provides compelling evidence for the presence of SP stars also along the cluster AGB (see Figs.~\ref{cubi} 
and~\ref{histograms}). Indeed, the presence of multiple stellar populations along the AGB, has been already detected photometrically in several GCs, using a similar approach as the 
one adopted here \citep[e.g.][]{sumo,miloneM2,milone2808,nardiello15}..

We also show that three stars analysed in \citet{marino08} should be classified as candidate AGB rather than RGB stars according to their position in the CMD. Such stars display a spread in 
[Na/Fe] $\sim$ 0.45 dex; i.e., they belong to different populations.

Our result is in apparent contrast with what found by \citet{maclean} spectroscopic study. 
An investigation of the reasons of this discrepancy is beyond the purposes of this study. However, we note that those authors admittedly expressed some caution about their results, i.e. 
their uncertainties in Na and O abundances did not allow to draw firm conclusions on the complete absence of SP stars in M~4 AGB.

Nonetheless, blue HB stars with the highest Na enrichment have the same maximum Na enrichment displayed by RGB stars \citep{mvm, villanova} and are expected to evolve to the AGB according to 
standard stellar evolution. Hence, the photometric discovery of SP stars in the AGB of M~4 well fits within the standard stellar evolution framework.

\section*{acknowledgements}
We thank the anonymous referee for a detailed report that helped to improve the presentation of our results.
CL gratefully acknowledges financial support from the European Research Council (ERC-CoG-646928, Multi-Pop, PI: N.~Bastian).
SC acknowledges the financial support by PRIN-INAF2014 (PI: S. Cassisi).

\bibliographystyle{mnras}
\bibliography{M4agbbib}

\bsp	
\label{lastpage}
\end{document}